# Old Story New Tell: The Graphite to Diamond Transition Revisited

Sheng-cai Zhu[1,2,&], Xiao-zhi Yan[3,&], Jin Liu[4,&], Artem R. Oganov[5], and Qiang Zhu[1]*

1 Department of Physics and Astronomy, University of Nevada, Las Vegas, NV 89154, USA

2 School of Materials, Sun Yat-sen University, Guangzhou 510275, China

3 Academy for Advanced Interdisciplinary Studies, and Department of Physics, Southern University of Science and Technology, Shenzhen, 518055, China

4 School of Mechanical Engineering, Jingchu University of Technology, Jingmen 44800, People's Republic of China

5 Skolkovo Institute of Science and Technology, 3 Nobel Street, Moscow 143026, Russia

**Abstract**

Graphite and diamond are two well-known allotropes of carbon with distinct physical properties due to different atomic connectivity. Graphite has a layered structure in which the honeycomb carbon sheets can easily glide, while atoms in diamond are strongly bonded in all three dimensions. The transition from graphite to diamond has been a central subject in physical science. One way to turn graphite into diamond is to apply the high pressure and high temperature (HPHT) conditions. However, atomistic mechanism of this transition is still under debate. From a series of large-scale molecular dynamics (MD) simulations, we report a mechanism that the diamond nuclei originate at the graphite grain boundaries and propagate in two preferred directions. In addition to the widely accepted [001] direction, we found that the growth along [120] direction of graphite is even faster. In this scenario, cubic diamond (CD) is the kinetically favorable product, while hexagonal diamond (HD) would appear as minor amounts of twinning structures in two main directions. Following the crystallographic orientation relationship, the coherent interface t-(100)gr//(11-1)cd + [010]gr//[1-10]cd was also confirmed by high-resolution transmission electron microscopy (HR-TEM) experiment. The proposed phase transition mechanism does not only reconcile the longstanding debate regarding the role of HD in graphite-diamond transition, but also yields the atomistic insight into microstructure engineering via controlled solid phase transition.

**1 Introduction**

Diamond, a stable form of carbon under high pressure, has been found naturally in the Earth's crust, as a result of carbon's evolution at HPHT conditions through the geological time scale[1–3]. Due to its many remarkable properties, diamond is great demand in both industry and fundamental research[4–7]. However, at ambient conditions graphite is more stable than diamond (with a difference of ~10 meV/atom from the modern quantum mechanical simulation based on density functional theory (DFT)[8] and ~17 meV/atom from more accurate diffusion Monte Carlo calculation[9]). Despite years of efforts, synthesizing diamond from graphite was not possible until 1950s[10,11]. Due to the distinct structural packing between graphite and diamond, the transition generally needs to go through a rather complicated pathway at HPHT conditions (15-18 GPa and 1500-2300 ℃)[5,12–15]. Fabricating diamond from other precursors, such amorphous carbon[16], carbon nanotubes[17], carbon nanoparticles[15] were considered as well. It was reported that controlling the microstructure of diamond, such as nano-twinning, is a key to promote the product's mechanical properties[18]. However, due to the lack of fundamental understanding of atomistic mechanism in these phase transitions, it remains challenging to realize a truly rational control of the product's microstructure during the process of synthesis.

Experiments show that cubic diamond (CD) is the main product under HPHT conditions[19,20]. Yet, another metastable form, hexagonal diamond (HD, also known as Lonsdaleite), was also observed in meteorites[21,22], shockwave experiments[23,24], as well as computer simulations[25]. Several experiments suggested that HD is the intermediate phase of graphite-to-diamond transition[26,27] based on a few newly observed X-ray diffraction (XRD) peaks. However, the assignment of new XRD peaks to HD is not well accepted due to the blurry nature of the pattern[28–31]. In a recent study, Németh et al reported that the hexagonal diamond cannot be obtained as a discrete material but only can be present as diamond {111} stacking fault or diamond (113) twins [32]. Thus, there is still no consensus that HD can be synthesized as a discrete material from the static compression of graphite. In addition to diffraction, electron microscopic techniques have been employed to study the graphite-diamond transition. A recent TEM experiment suggested that graphite transforms to diamond without any intermediate phases but through two coherent interfaces between graphite and CD, namely (100)gr//(11-1)cd and (001)gr//(111)cd[33]. Following the orientation relationship (OR), the coherent interface can be uniquely defined by the relation between specific planes and directions of two crystals on either side of boundary. Thus, they can be expressed by two parallel crystal planes $(hkl)_a//(h'k'l')_b$ and two parallel directions $[uvw]_a$ and $[u'v'w']_b$, where [uvw] and [u'v'w'] lie in the (hkl) and (h'k'l') planes. Wheeler[20] found that the shock-quenched diamond contains both CD and HD domains with orientations of (100)gr//(11-1)cd + [010]gr//[1-10]cd and (100)gr//(001)hd + [010]gr//[010]hd, respectively. They proposed that the phase transition can be achieved by the displacement of adjacent pairs of <-120> row with relative shears on alternating graphite basal planes. Despite these encouraging successes, collecting and interpreting the diffraction and TEM data under HPHT conditions remains challenging in general.

Complementary to the experimental studies, atomistic simulations can access a broader pressure-temperature space and provide insights into the phase transition at atomic level. In the past, several possible pathways have been proposed[25,34–36], including puckering, buckling and lateral displacement mechanisms[37–39] (as summarized in Figure 1). Among them, the puckering mechanism were suggested in many studies. For example, Fahy et al[34,35] proposed that graphite (001) plane would transform to the chair architecture of CD (111) or HD (001) plane under compression. In this mechanism, the interplanar distance in graphite first collapsed, leading to a puckering of the graphitic basal planes. Then, the puckered planes suddenly undergo an electronic reconfiguration from $sp^2$ to $sp^3$ state. The whole large graphite (001) plane puckers and transform to CD (111) or HD (001) layer-by-layer homogenously[40]. Similar to puckering, the buckling mechanism suggests that the graphite (001) plane will transform to the boat architecture of HD (100), and then complete the entire transition to HD. In a recent work, Xie et al[26] suggested that this mechanism yielded the lowest energy barrier pathway by the state-of-the-art transition state sampling method. Unlike the collective motion in either buckling or puckering, lateral displacement mechanism requires a group of carbon atoms belonging to adjacent graphite sheets stochastically vibrating out of graphite basal layer (up or up/down) and thus transform to CD or HD. Via the first-principles MD simulation, Scandolo[36] found that graphite transits to CD while HD exists as the twinning of CD. Interestingly, the preferred orientation relation from their MD simulation can be interpreted as (100)gr//(11-1)cd + [010]gr//[1-10]cd. Tateyama[40] also found that a similar mechanism is preferred when a large strain is allowed. Using an artificial neural-network (ANN) potential, Khaliullin et al[25] carried out a large-scale atomistic simulation to study the energetics of the nucleation mechanism of the graphite-diamond phase transition. In order to get the nuclei energies, they seeded diamond nuclei inside graphite matrix (~100×100×100 Å) and optimized the geometry by constant-pressure molecular dynamics simulations at 1,000 K for 30 ps to relax the atoms around the constrained region. Khaliullin's work provided an important first step to understand the nucleation mechanism of graphite-diamond transition in bulk at atomistic level. Yet, using the ANN potential to perform a HPHT MD simulation on a large cell (~10-50 nm) at the timescale of nano seconds is still beyond our computationally cost. Therefore, the mechanism of graphite-diamond phase transition at the realistic HPHT conditions remains elusive.

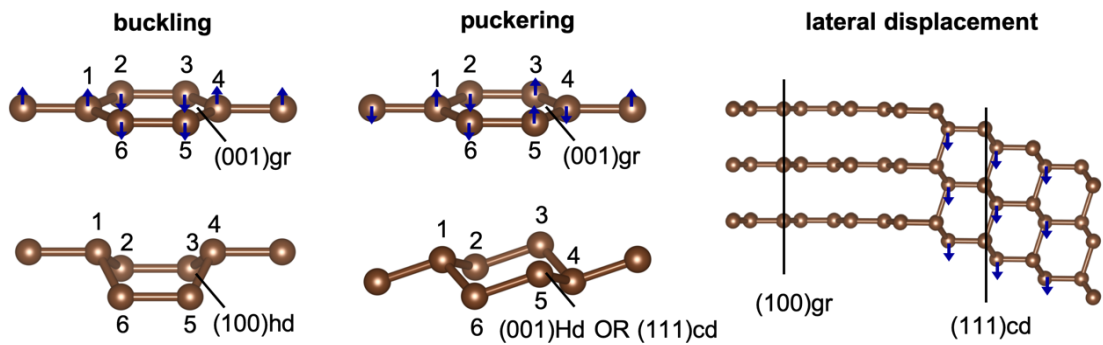

**Figure 1. Atomistic mechanisms of graphite to diamond transition.** (a) buckling, (b) puckering and (c) lateral displacement.

To our knowledge, most of the theoretical results were from the simulations based on small structural models under the assumption of homogeneous nucleation. When using the periodic boundary conditions, the entire (001)gr basal layer will turn into (111)cd, (001)hd or (100)hd. However, the minimum number of atoms required to trigger such a transition (nuclei core) is unknown. Therefore, the calculated energy barriers for a periodic unit cell may not be instructive in evaluating the kinetic preference of different transition pathways under consideration. For instance, HD formation was calculated to be preferred over CD formation due to such energy barrier comparison[26], which is contrary to the experimental observation. To our knowledge, the current state of the art to determining the energy barrier based on DFT is reliable, but the small periodic unit cell may not be suitable for describing the entire graphite-to-diamond phase transition since graphite is a system with strongly anisotropic behavior. To account for such effect, a large simulation model is needed, while most of the previous studies have been restricted to simulation cells with only a few tens of atoms. Last, in a realistic model, the graphite sheet is supposed to depart from the equilibrium state due to thermal fluctuation under finite temperature and local stress. Therefore, these previously reported mechanisms may be limited in describing the transitions occurring at HPHT conditions.

To fully understand this underlying phase transition mechanism, we conducted a complete study at different scales. First, we revisited the most likely graphite/diamond interface models under compression through an exhaustive sampling of small periodic unit cells. Second, we carried out a series of large-scale MD simulations based on a newly developed angular dependent potential (ADP) to gain an atomistic understanding of the diamond nucleation and growth under HPHT conditions. Our MD simulations demonstrated that the nuclei emerge at the graphite grain boundaries. Strikingly, we found that the commonly believed [001]gr direction is not the only preferred growth direction, instead, the propagation of diamond along [120]gr with the crystallographic orientation t-(100)gr//(11-1)cd + [010]gr//[1-10]cd is much faster. Following this mechanism, CD is the main product while HD can exist as the twinning structure. The coherent interface resolved from high resolution TEM images is consistent with our MD simulation. Findings from this study shed light on the longstanding debates of graphite-to-diamond phase transition, and facilitate understanding of the anisotropic behavior of the (001)gr plane under HPHT conditions. In addition, we propose a route to fabricate superhard diamond by harvesting the twin structures along [120]gr at the microstructural level via the *pre-bent* graphite sheets.

## 2 Results

**2.1 Transition State Sampling at static conditions.**

We started our investigation by scanning the low-energy intermediate interface models between graphite and diamond. To explore them exhaustively, more than 10,000 energy minima were visited by stochastic surface walking (SSW) method[41–43] together with the high-dimensional neural networks (NN) potential[44]. The sizes used in the simulation models range from 12 to 126 atoms per unit cell. Among them, seven lowest interface energy structures between graphite/CD and graphite/HD were extracted for further analysis in detail (see Figure 2). They were named according to the transition products with different interfacial energy, namely GH1 (0.39 eV/Å$^2$), GH2 (0.19 eV/Å$^2$), GH3 (0.31 eV/Å$^2$), GH4 (0.21 eV/Å$^2$), GC1 (0.37 eV/Å$^2$), GC2 (0.19 eV/Å$^2$), and GC3 (0.36 eV/Å$^2$). By inspecting their geometries, we found that alignment of HD domain in GH1 is 90 ° off relative to that in GH2, GC2 is 54.75 ° off relative to GC3, while GH2 and GH3 have the same crystal orientation but different coherent plane.

Clearly, all three previously proposed mechanisms have been covered by these interface models. For instance, GH1 and GC1 follow the puckering mechanism where the flat (001)gr transforms to chair architecture of HD and CD, respectively; while the GH3 follows the buckling mechanism where (001)gr transforms to the boat architecture of (100)hd plane. In these models, diamond is supposed to grow along the direction perpendicular to graphite sheets. On the other hand, GH3, GH4, GC2 and GC3 follow the lateral displacement mechanism. In detail, GH3, GC2 and GC3 can be obtained by carbon atoms vibrating out of the (001)gr plane of graphite in [010]gr view while GH4 is in [1-10]gr view. Since the CD (111)cd plane and HD (001)hd plane have similar transition mechanism and atomic arrangements, they can co-exist and form a hybrid interface, namely GH3/GC3, as Xie et al. reported[26]. Our results are largely consistent with those structures found in Xie et al, except that we found two new interface structures between graphite and CD, namely GC2 ((100)gr//(010)cd+[010]gr//[101]cd) and GC3 ((100)gr//(11-1)cd+[010]gr//[1-10]cd). It should be noted that GC3 is very similar to the fragment of a mixed phase (GCH) proposed by Xie et al.[26] However, the authors failed to provide the entire GC3 model due to the limit of simulation model size. Indeed, the GC3 interface structure was observed by Wheeler et al[20] in their shockwave experiment. This also marks the importance of performing the simulation with a sufficiently large model.

For the suggested pathways, it is important to investigate their energy barriers during the graphite-diamond transition. The average energy barriers of the puckering mechanism GH1 and GC1 are 0.31 eV/Å$^2$ and 0.28 eV/Å$^2$, respectively. GH2 has the lowest barrier, 0.16 eV/Å$^2$. For the lateral displacement mechanism (GH4 and GC2), carbon atoms just need to shift by a 1/2 of the inter-layer with the energy barriers 0.22 eV/Å$^2$ and 0.19 eV/Å$^2$, while GH3 and GC3 need to overcome higher energy barriers (0.29 eV/Å$^2$ and 0.35 eV/Å$^2$) by shifting 1/3 of the inter-layer.

| Name | Interface structure | OR | Ea | γ | Name | Interface structure | OR | Ea | γ |
|---|---|---|---|---|---|---|---|---|---|
| GH1 | | $(001)_{gr}//(001)_{hd}$ $[010]_{gr}//[010]_{hd}$ | 0.31 | 0.39 | GC1 | | $(001)_{gr}//(111)_{cd}$ $[010]_{gr}//[1\text{-}10]_{cd}$ | 0.28 | 0.37 |
| GH2 | | $(001)_{gr}//(100)_{hd}$ $[010]_{gr}//[010]_{hd}$ | 0.16 | 0.19 | GC2 | | $(100)_{gr}//(010)_{cd}$ $[010]_{gr}//[101]_{cd}$ | 0.19 | 0.19 |
| GH3 | | $(100)_{gr}//(001)_{hd}$ $[010]_{gr}//[010]_{hd}$ | 0.29 | 0.31 | GC3 | | $(100)_{gr}//(11\text{-}1)_{cd}$ $[010]_{gr}//[1\text{-}10]_{cd}$ | 0.35 | 0.36 |
| GH4 | | $(110)_{gr}//(100)_{hd}$ $[1\text{-}10]_{gr}//[001]_{hd}$ | 0.22 | 0.21 | | | | | |

**Figure 2. The phase transition path of graphite to diamond under small unit cell, interface, orientation relationship, energy barrier (eV/Å$^2$) and interface energy (eV/Å$^2$).** The propagation directions are marked by the blue arrows.

**2.2 Direct modeling of the phase transition from large-scale MD simulations.**

These atomic interface models are instructive to understand the possible atomistic mechanism of the graphite-diamond transition. However, they are limited by the size of simulation model and cannot describe the phase transition under realistic conditions. To overcome these limits, we employed the large-scale MD simulations to directly study this phase transition at the relevant HPHT conditions. In this study, both the single- and poly-crystalline graphite models were considered. Simulations were carried out in the NPT ensemble with incremental pressures. MD simulations probe only sufficiently fast and frequent processes. First-order phase transitions like graphite to diamond, typically have a high activation barrier, and can only be seen by MD at pressures or temperatures exceeding those of equilibrium phase transition.

Therefore, we used the exceeding pressure to accelerate the phase transition in our MD simulation. To start, both the single-crystal system (150,000 atoms) and polycrystal system (1,226,000 atoms, composed of several grains with randomly generated orientations), were firstly relaxed at 25 GPa and 1500 K with sufficiently long equilibration time of 0.20 nano seconds (ns). Then, the pressure was steadily increased until the phase transition was observed. After a few test runs, we optimized the final pressures to be 40 GPa for polycrystal and 80 GPa for single crystal with compression rate 5 GPa/ns and 16.66 GPa/ns, which allowed us to study the phase transition at lower critical pressure conditions with affordable simulation cost.

A typical MD trajectory of polycrystalline graphite under compression is recorded in animation 1 (supporting information) and depicted in Figure 3. Clearly, the nuclei were initiated by the local (001)gr plane distortion at the graphite's grain boundaries, followed by the propagation along different directions. We can better understand the entire trajectory from the evolution of several key thermodynamic quantities such as energy and volume (Figure 3a). The entire process can be split into three stages. At the first stage (<0.9 ns), the energy (volume) of the system smoothly increased (decreased). When it reaches a critical pressure (29 GPa), there is a dramatic change of the slope, signaling the formation of diamond nuclei and their propagation in the graphite matrix (0.9-1.6 ns). Finally, the system reaches a stage without significant structural change (>1.6 ns). Note that there still exists a notable fraction of untransformed graphite in the system, in agreement with previous experimental observations that the graphite-diamond transition could not complete in a short time period.

We also analyzed the statistics in each MD snapshot. As shown in Figure 3b, we found that the ratio of carbon atoms in graphite pattern steadily decreased while the ratio of CD atoms increases with time. Interestingly, we also observed a certain ratio of atoms labelled as HD, which appeared after the formation of nucleation of CD (0.91 ns versus 1.21 ns). Combining the detailed image analysis (Figure 3d), we found that HD exists as twinning structures (fewer than 4 layers) which were randomly located in the CD matrix. Thus, HD should be better interpreted as a byproduct during the growth of CD in the graphite matrix. Though several previous experiments suggested that HD is the intermediate phase of graphite-to-diamond transition[26,27] based on a few newly observed X-ray diffraction (XRD) peaks, the assignment of new XRD peaks to HD is still controversy. From a thorough HRTEM experimental analysis, Németh et al suggested that HD exists as twinning structures[32]. Here, our MD simulation provides the direct theoretical evidence to support Németh's observation.

Obviously, the growth of diamond nuclei exhibits a strong anisotropic behavior. At the beginning, the shapes of nuclei were nearly sphere-like (see Figure 3d). Later, we found that diamond propagated quickly along the [120]gr direction after the nuclei formation while the growth along [001]gr direction is slower, see Figure 3c. We also selected three different diamond nuclei (marked as A, B, C in Figure 3d) from the simulation and monitored their growth as a function of time. In general, the growth rate of (100)gr//(11-1)cd is about 2.5 times that of (001)gr//(111)cd. This is contrary to the previous results based on the small simulation models at zero temperature[25,34,35] which suggests that the preferred growth direction is either [001]gr or [120]gr. The high strain graphite on (100)gr//(11-1)cd interface with smaller d spacing is unstable, that's why the propagation along [120]gr direction is much faster. Therefore, the entire transition should follow a hybrid mechanism involving both growth at both lateral and normal directions of the graphite sheets.

Our simulation also differs from the previous DFT simulations in terms of the interface geometry. In the MD simulation, we observed that the graphite (001) in the interface structure (Figure 3d) is slightly tilted compared to the ideal (001) plane. This leads to a non-orthogonal dihedral angle of 70.5 ° between the tilted graphite (001) and diamond (11-1) planes, which agrees well with the experimental observation (to be discussed in the following section). Therefore, we name this interface as t-(100)gr//(11-1)cd + [010]gr//[1-10]cd. Note that the tilted graphite (001) plane will return to the ideal alignment under the geometry optimization in both DFT and force field calculations at 0 K. It also should be mentioned that the general lateral displacement mechanism allows the formation of phase growth propagation frontiers with high index planes. If the diamond nuclei propagate along [120]gr with different rates, different phase growth propagation frontiers will occur, as shown in Figure 4e and Figure S2. In this case, other high index twins or stacking faults may occur when these high index planes meet highly bent graphite during the phase propagation. On the other hand, under high temperature, strong thermal fluctuation can introduce a pronounced corrugation to the planar graphite sheet and thus change the geometry significantly. Therefore, a large-scale modeling at high temperature is necessary to describe the entire transition at the atomic level.

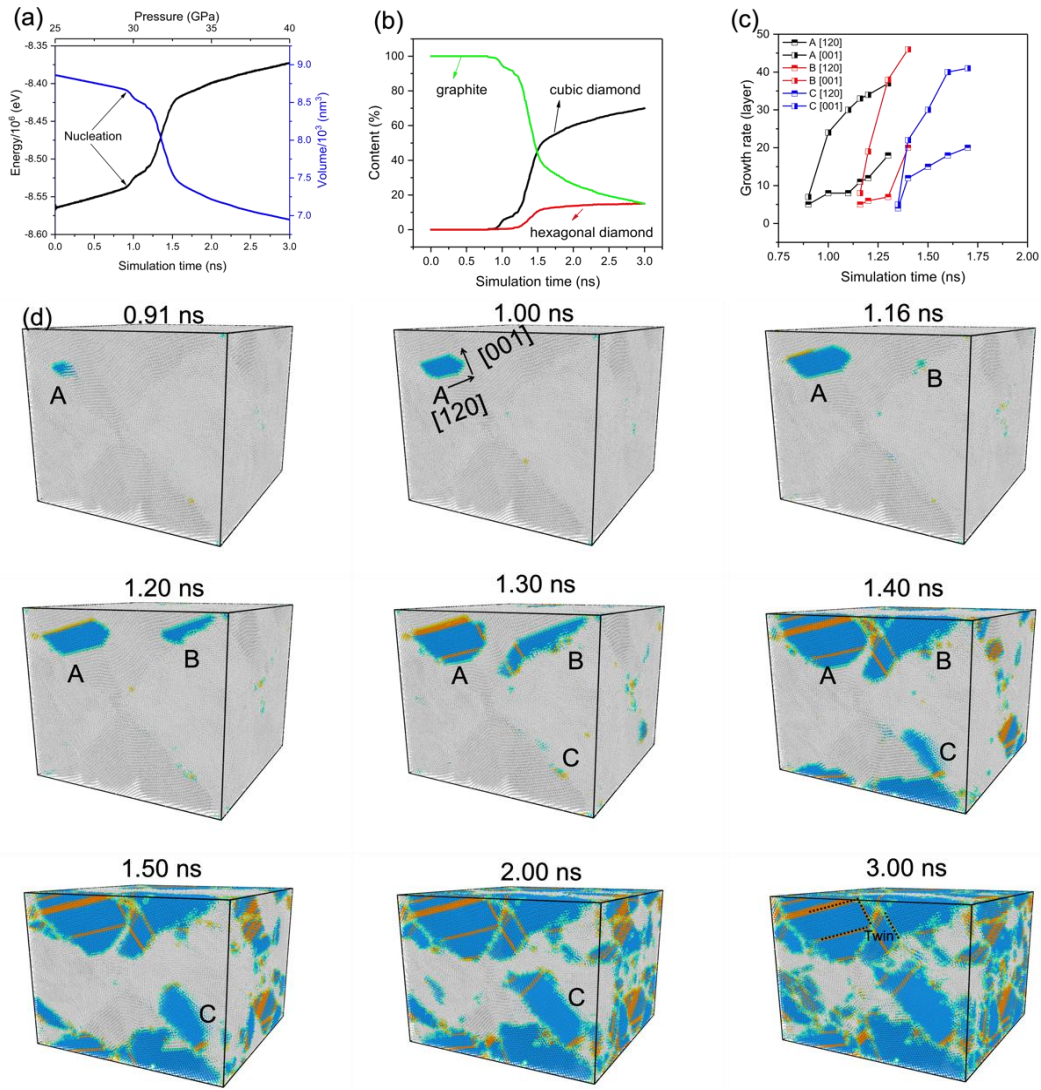

**Figure 3. Statistical analysis of MD simulation.** (a) The energy and volume curves as a function of the simulation time. (b) The growth rate of graphite, cubic diamond and hexagonal diamond during the simulation time. (c) The anisotropic phase transition behavior of A, B and C domain in [120]gr and [001]gr directions. (d) List of representative snapshots at different intermediate stages of simulation. In the snapshots, the atoms representing graphite, cubic diamond and hexagonal diamond are marked by grey, blue and orange spheres, respectively.

We also repeated the MD calculation with faster compression rate for both polycrystalline (poly 2: 10 GPa/ns) and single crystalline (single: 16.66 GPa/ns) samples. For the polycrystalline sample with faster compression rate, no significant differences in the transformation path were observed except that the transition pressures were slightly shifted to a higher value. For the single crystals, the critical pressure is much higher (60 GPa), which is expected since the single crystal has no defects to facilitate nucleation.

**2.3 Experimental verification**

To verify the simulation results, we conducted a series of experiments. Polycrystalline diamond was synthesized at HPHT and processed by focused ion beam (FIB). Further, the samples were analyzed by TEM. From the low-magnification TEM image, we can find that the main product is CD, which is consistent with our simulation (see Fig 4a and b). Also, the dark field TEM and the HRTEM image suggests that HD is the stacking fault (twinning) structure of CD (see Fig 4b and c). The high-resolution TEM images of the sample in Figure 4d can easily distinguish the mixture of graphite and diamond. Specifically, the observed d-spacings of 3.35 and 2.07 Å are graphite (001) plane and diamond (111) plane, respectively. The dihedral angles between the graphite t-(001) and diamond (11-1) in the interface of t-(100)gr//(11-1)cd + [010]gr//[1-10]cd is 71($\pm$1)°. Considering the fluctuation of graphite (001) plane, the measured dihedral angle is very close to the results from our MD simulation (~70°, see the interface model in Figure 4e and f. Though the (001)gr//(111)cd + [010]gr//[1-10]cd interface has been observed often, the interface on (100)gr plane (Figure 4g and h) was rarely reported[29,33]. This could probably be explained by the fact that the transition along the [120]gr direction is much faster and thus less likely to be captured by ex-situ experiment. The identification of the t-(100)gr//(11-1)cd + [010]gr//[1-10]cd interface is essential for us to understand the complete picture for this transition.

Besides the HRTEM from our own experiment, two main interface models, defined as the GC1 (001)gr//(111)cd and t-GC3 (100)gr//(111)cd, were reported in the previous literature [29, 33] as well (see Figures 4g and 4h). For example, viewing along [1-10]cd direction, two {111} diamond fringes can be found on Figures 4g and 4h, and the (100)gr is coherent to (111)cd with a dihedral angle of ~71($\pm$1)° between (001)gr and (111)cd. Moreover, the GC1 interfaces were found six times, while there exists only one GC3 interface structure in Ref. [33]. This can be explained by the fact that the growth along [120]gr is much faster than [001]gr. Once (100)gr//(11-1)cd interfaces formed, they quickly propagate along [120]gr and eventually transform to diamond. The relative occurrences of two interface structures support the identified preferred propagate directions observed in our MD simulation.

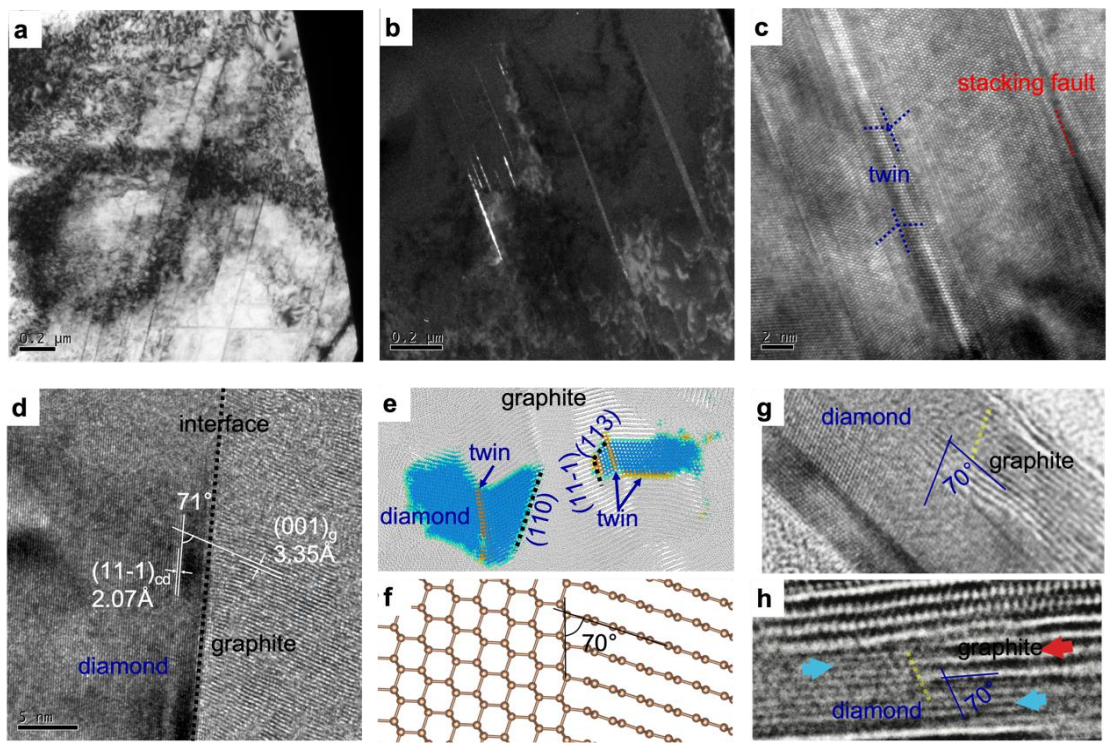

**Figure 4. TEM image analysis.** (a−b) The cubic diamond TEM, inset of (a) is the selected area electron diffraction (SAED), (c) The highlighted twin structure. (d) The HRTEM of phase junction between graphite and diamond. (e) The phase junction slab from MD simulation of the polycrystal system. (f) The junction from the small cell simulation. (g-h) The experiment phase junction from ref. [29] and ref. [33] both with t-(100)gr//(11-1)cd + [010]gr//[1-10]cd and (001)gr//(111)cd + [010]gr//[1-10]cd interface (Copyright 2001, with permission from xxx). The yellow dotted lines in g and h were drawn to emphasize the region of graphite-diamond phase boundary.

## 3 Discussions

In the past, the graphite-to-diamond phase transition was mostly suggested to follow either puckering or buckling mechanism, where both prefer the growth of CD/HD from graphite in a layer-by-layer manner along the [001]gr direction. However, both our MD simulation and TEM analysis suggested that there should exist two preferred crystal growth directions, i.e, [120]gr and [001]gr. In fact, the growth along [120]gr is even faster than along [001]gr. This can be understood by the classical nucleation theory[37,45], which expresses the thermodynamic potential change ($\Delta G$) on forming the nucleus as

$$\Delta G = n[\Delta g(p,T) + E_\varepsilon] + \gamma S \quad (1)$$

where *n* is the number of molecules in the nucleus, $\Delta g(p,T) = g_\beta - g_\alpha$, is the difference in specific thermodynamic potentials (per molecule) between the initial ($\alpha$) and the new ($\beta$) phases, $E_\varepsilon$ is the strain energy (per molecule) in the matrix-nucleus system after nucleation, $\gamma$ is the specific energy of the interphase boundary, S is the interfacial area. The kinetic preference can be quantified by $\frac{\partial \Delta G}{\partial n}$. Since (100)gr//(11-1)cd and (001)gr//(111)cd have a similar specific interface energy ($\gamma$, see the DFT results), the nucleation growth is dominated by strain energy term ($E_\varepsilon$). To obtain a quantitative understanding, we selected two typical interface models GC1 and GC3 with half diamond and half graphite in the superlattice. If diamond grows along [001]gr, the penalty energy $\frac{\partial \Delta G}{\partial n}$ due to lattice distortion can be described by the strain energy of GC3. On the other hand, the strain energy of GC1 denotes the corresponding $\frac{\partial \Delta G}{\partial n}$ for the growth along [120]gr (see Figure 5d and e). According to our calculation:

$$E_\varepsilon = \frac{[E_{\text{distorted}}(\text{gr}) - E_{\text{ideal}}(\text{gr})] + [E_{\text{distorted}}(\text{CD}) - E_{\text{ideal}}(\text{CD})]}{N} \qquad (2)$$

For the 128-atoms GC1 model the total strain energy is merely 9 meV/atom, and the 32-atoms GC2 total strain energy is 93 meV/atom, whereas the strain energy is 184 meV/atom for the 96-atoms GC3 model. It is important to note that the models of GC1 and GC3 only account for two special cases where there exist half diamond and half graphite. Hence, the calculated $E_\varepsilon$ values and propagation rate will vary during the growth of diamond. From our MD simulation, we found the overall growth along [120]gr is 2.5 times faster than [001]gr. In real experiments, the ratio may be even larger since t-(100)gr//(11-1)cd + [010]gr//[1-10]cd interface was seldom identified. Despite this numerical variation, we can safely conclude that the [120]gr growth is kinetically more favorable than [001]gr growth due to a notable difference between the strain energies.

The mechanism can also reconcile the long debates on the role of HD in the graphite-diamond transition. In our simulation, we exclude the possibility of HD as the intermediate phase between graphite-to-diamond transition at HPHT conditions. HD can exist at twin boundaries of CD during the growth of CD nuclei. It is interesting to probe the formation of twins along both directions. When two CD nuclei meet along the [001]gr direction, they would stochastically form either stacking fault, a twin boundary or perfect conjunction. On the other hand, the formation of twins along [120]gr largely depends on the local corrugation of graphite sheets involved in the transition. As shown in Figure 4e, the twins tend to appear where the graphite sheets are locally bent. When graphite sheets are bent, forming the twin structure is apparently the best way to minimize the total penalty energy due to lattice mismatch. Thus, introducing the corrugation to the graphite sheet is a key to produce the twins. In our MD simulation, the HPHT conditions, together with the local structural defects (i.e., grain boundaries), provide the sources to bend the graphite sheets and thus produce the twins along [120]gr. Similar results were also found in several other experiments [29,33] and MD simulations[46]. In particular, a recent experiment[18], through starting from another precursor of carbon onion nanoparticles, also reported the presence of many twins in two different directions (see Figure S3 in SI). This can be well understood by the fact that the graphite sheets in the carbon nanoparticles have been *pre-bent* to adopt the onion like arrangement[18]. Note that it is much harder to produce the [120]gr direction twins if the graphite sheets adopt a purely planar configuration. For instance, many synthetic diamonds from single crystal graphite were reported to have the lamellar texture with only one direction of twin structures along [001]gr direction in a nano domain[5,45], see Figure S4. As a result, the carbon nanoparticles with *pre-bent microstructure* can enhance the production of twins along the [120]gr direction, which thus promote the hardness of the synthetic diamonds.

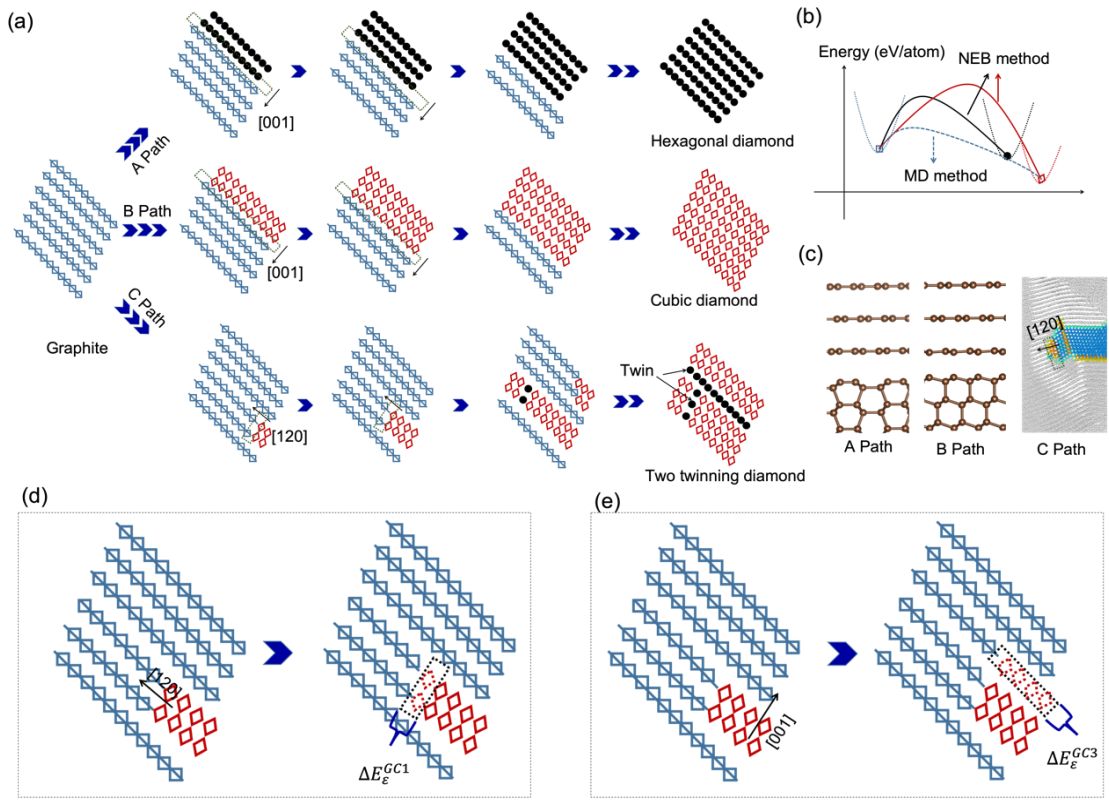

**Figure 5. Schematic phase transition mechanisms.** (a) Schematic representation of nucleation and growth of the graphite to CD and HD, including graphite to HD transition and growth in [001]gr direction (**A path**, top), graphite to CD transition and growth in [001]gr direction (**B path**, middle), graphite to CD transition with HD along both [120]gr and [001] directions (**C path**, bottom). (b) The schematic energy barrier comparison for different paths (A: black; B: red; C: dot). (c) The interfaces constructed from different paths. (d) The penalty strain energy for [120]gr direction growth $\Delta E_\varepsilon^{GC1}$. (d) The penalty strain energy for [001]gr direction growth $\Delta E_\varepsilon^{GC3}$.

Let us summarize our findings for the graphite-to-diamond phase transition under HPHT conditions. While previous studies suggested the growth of diamond has only one preferred direction [001]gr (paths A and B in Figure 5a), our results showed that graphite transforms to CD with two preferred growth directions ([001]gr and [120]gr, path C in Figure 5a), and [120]gr is more favorable. The defects developed at graphite's grain boundaries help to trigger the formation of CD nuclei. The growth along [120]gr is generally faster than [001]gr due to the anisotropic strain distribution at the graphite/CD interfaces. Following this mechanism, the HD will appear when two CD nuclei meet in these two main growth directions. In particular, the occurrence of diamond twins along [120]gr is determined by the local configuration of graphite sheets involved in the phase transition. Therefore, the mechanism developed here can be used to tailor the mechanical properties of diamond by controlling its microstructure during the synthesis.

## 5. Conclusion

In summary, combining both large-scale MD simulation and HRTEM measurement, we propose a new graphite-to-diamond phase transition mechanism that can resolve several longstanding issues. We found that the graphite-to-diamond transition at HPHT conditions involves two preferred growth directions, in which the growth of nuclei along [120]gr direction is faster than that along [001]gr direction due to an anisotropic distribution of interfacial strains. The coherent interface orientation relation resolved from HRTEM, t-(100)gr//(11-1)cd + [010]gr//[1-10]cd, confirmed the growth along [120]gr observed in our MD simulation. Following this mechanism, CD is the main product while HD is present as the twin boundary. This mechanism is also supported by previous data where two interface structures were found and the observation of twin structures in two different directions ([120]gr and [001]gr). The results of this work rationalize that the graphite to diamond is largely rooted in the anisotropic behavior of the (001) plane and also suggest a route to fabricate the twin structures along [120]gr by pre-bending the graphite sheets. Understanding this mechanism can help better engineer the microstructure of synthetic diamonds from HPHT conditions, which has been demonstrated to have a great impact on mechanical properties such as hardness of the resulting diamond.

**Theoretical and Experimental Methods**

**Reaction Pathway Sampling**

In order to explore the potential energy surface (PES) of carbon under high pressure, we employed the recently developed SSW method which integrated with first principles DFT method (SSW-DFT)[41–43] and the high dimensional neural networks (NN) potential[44] to sample the low energy interfaces and pathways. The SSW reaction pathway sampling is based on SSW global optimization method which is able to explore complex PES to simultaneously identify both structures and reaction pathways. For solid phase transitions, this is to identify the one-to-one correspondence for lattice (L($e_1$,$e_2$,$e_3$), $e_i$ being the lattice vector) and atom ($q_i$, i=1,..3N, N is the number of atom in cell) from one crystal phase (the initial state, IS) to another (the final state, FS), which constitutes the reaction coordinates of the reaction, i.e. $Q_{IS}(L,q) \rightarrow Q_{FS}(L,q)$. In one SSW pathway sampling simulation, we need to collect as many as possible IS/FS pairs (typically a few hundreds) to ensure the identification of the best reaction coordinate, the one corresponding to the lowest energy pathway. With such a pair of reaction coordinates, $Q_{IS}(L,q)$ and $Q_{FS}(L,q)$, it is then possible to utilize variable-cell double-ended surface walking (VC-DESW) method[43] to identify the reaction transition state (TS) and the minimum energy pathway.

The SSW pathway sampling is fully automated and divided into three stages in simulation, namely, (i) pathway collection via extensive SSW global search; (ii) pathway screening via fast DESW pathway building; (iii) lowest energy pathway determination via DESW TS search. The first stage is the most important and most time-consuming part, which generates all the likely pairs of generalized reaction coordinates linking different crystal phases. For the carbon phase transition in this work, we have collected more than 1000 pairs $Q_\alpha(L,q)$ and $Q_\beta(L,q)$, which leads to the finding of the lowest energy pathway. The lowest energy pathway obtained from sampling was then analyzed to identify the key atom displacement patterns. Then, using the information, we further enlarged the supercell up to 60~120-atom per cell via the interface intermediate structure mechanism (dependence on the interface structure) and re-searched the lowest energy pathway, which is found to dramatically lower the overall reaction barrier. The stability of the interfaces was evaluated by considering the interfacial energy[48], defined as $\gamma=(E_{tot}-E_a-E_b)/2S$, where $S$ is the interfacial area, $E_a$ and $E_b$ are the energies of the parent phases, and $E_{tot}$ is the energy of the mixed phase.

For each image generated from SSW sampling, the energy and forces were calculated by the plane wave DFT program Vienna *ab initio* simulation package (VASP)[49]. The electron-ion interaction of C atoms was represented by the projector augmented wave (PAW)[50] scheme and the exchange-correlation functional utilized was GGA-PBE[51]. For all the structures, both lattice and atomic positions were fully optimized in SSW until the maximal stress component is less than 0.1 GPa and the maximal force component is less than 0.01 eV/Å.

**Molecular Dynamics Simulation**

To enable the large-scale MD simulation, an interatomic potential was developed for elemental carbon. Since the phase transition involves the break and formation of covalent bonds, an embedded atom model (EAM) formalism with angular dependent potential (ADP) was employed to fit the potential energy landscape of carbon based on massive DFT data. The detail of the carbon ADP potential will be published elsewhere. In order to validate the accuracy of the ADP potential, we calculated the equation of states for both graphite and diamond, and compared them with the results from DFT. In addition, the energy barriers calculated by DFT and the ADP potential are consistent with each other, which indicate the accuracy of ADP potential is sufficient for the purpose of this study. More details can be found in the supporting materials.

All MD simulations were run in the LAMMPS code[52]. In our calculation, four different initial models (one single crystalline and two polycrystalline graphite models) were used. For the single crystalline sample, there are 150,000 carbon atoms of graphite structure in a box with a = 11 nm, b = 12 nm and c = 10 nm. In the bigger polycrystalline graphite model, we generated 4 grains randomly orientated in a ~23.4 nm cubic box with total ~ 1226,000 atoms. In the smaller polycrystalline model, about 20 small grains were randomly orientated in a ~20 nm cubic box with total ~840,000 atoms. All initial geometries were heated to 1500 K at ambient pressure condition for an equilibration of 0.2 ns. Then, the constantly increasing pressures with a damping value of 1.0 were applied to these samples for 3.0 ns. To understand the process, we choose to output the enthalpy, pressure, volume and atomic structures every 1 ps. Simulations were carried out in the NPT ensemble, and the thermostat was employed to maintain a constant temperature. The simulation results were visualized and analyzed in the OVITO package[53].

**Experimental synthesis and characterization**

Polycrystalline diamond was synthesized from graphite under HPHT conditions. Graphite was pressed into a pellet, and then processed in a two-stage multi-anvil apparatus based on a DS6 x 25 MN cubic press machine. Samples were compressed to the desired values before heating and then heat 1000-2000 °C for 0.5-2 h at 14GPa. The pressure was estimated by the well-known pressure-induced phase transitions of Bi, ZnTe, and ZnS. The treating temperature was directly measured by using W97Re3-W75Re25 thermocouples. After cooled to room temperature, the samples were prepared by focused ion beam (FIB) and characterized with high resolution TEM (FEI Tecnai G2F20 S-Twin, USA, operated at 200 kV).

**ASSOCIATED CONTENT**

Supporting Information

The animations of poly-crystal molecular dynamic simulation (Animations 1, 2); The animation of single crystal molecular dynamic simulation (Animation 3); The validation of carbon potentials (Tables S1 and Figure S1 and S2); The schematic phase prorogation along high index diamond planes; The high resolution TEM images of two twinning directions structure from literature (Figure S3); The lamellar texture cubic diamond (Figure S4) are in Supporting Information.

The Supporting Information is available free of charge on http://www.nature.com**.**

**AUTHOR INFORMATION**

Corresponding Author

*Qiang Zhu (qiang.zhu@unlv.edu)

Author Contributions

& These authors contributed equally: Sheng-cai Zhu, Xiao-zhi Yan and Jin Liu.

**Notes**

The authors declare no competing financial interest

**Acknowledgement**

We thank Hong-wei Sheng for providing the carbon-ADP potential. Work at UNLV is supported by the National Nuclear Security Administration under the Stewardship Science Academic Alliances program through DOE Cooperative Agreement DE-NA0001982. We acknowledge the use of computing resources from XSEDE (TG-DMR180040). Sheng-cai Zhu and Xiao-zhi Yan are supported by NSFC (Grant No: 21703004 and 11704014).

**Support information**

1. **The C-ADP potential**

A general-purpose interatomic potential has been developed for carbon simulation. The potential is based on the angular-dependent-potential (ADP) formalism [1], which is an extension of the embedded atom method (EAM). [2] The angular dependence of the interatomic interactions is treated by the multipole expansion approach up to the quadrupole term. The ADP formalism is expressed as:

$$E_{total} = \frac{1}{2}\sum_{i,j} \Phi(r_{ij}) + \sum_i F_i(\rho_i) + \frac{1}{2}\sum_{i,\alpha} \mu_{i\alpha}^2 + \frac{1}{2}\sum_{i,\alpha,\beta} \lambda_{i\alpha\beta}^2 - \frac{1}{6}\sum_i v_i^2$$

where $i, j$ enumerate atoms and the superscripts $\alpha, \beta = 1,2,3$ refer to the Cartesian directions. The first two terms are the classical treatment of the EAM potential, where $\Phi(r_{ij})$ is the pair potential, and $F_i(\rho_i)$ is a functional to denote the contribution due to electron density $\rho_i$ of each atom. The additional three terms introduce non-central components of bonding through the vectors

$$\mu_{i\alpha} = \sum_{j \neq i} u(r_{ij}) r_{ij\alpha}$$

and tensors

$$\lambda_{i\alpha\beta} = \sum_{j \neq i} w(r_{ij}) r_{ij\alpha} r_{ij\beta}$$

$v_i$ are traces of the $\lambda$-tensor:

$$v_i = \sum_\alpha \lambda_{i\alpha\alpha}$$

These additional terms can be thought of as measures of the dipole ($\mu$) and quadrupole ($\lambda$) distortions of the local environment of an atom.

In this work, the five functions, $\Phi(r)$, $F(\rho)$, $\rho(r)$, $u(r)$ and $w(r)$, represented with spline interpolations, were parametrized based on ab initio calculations. We used the potfit code[3,4] for parametrization. The ab initio training dataset contains ~ 4000 atomic configurations (including graphite, cubic/hexagonal diamond and many other known structures, as well as configurations from NPT-MD simulation at a series of temperatures) and the total number of atoms in the potential fitting reach as many as $3 \times 10^6$. More details of the potential development will be presented elsewhere, and the C-ADP potential is available upon request.

## 2. Validation of the C-ADP potential

In this work, we developed a new carbon angular dependent potential (ADP) based on the embedded atom model (EAM) formalism to conduct the large-scale molecular dynamics (MD) simulation. The choice of ADP is based on a survey of several popular carbon potentials in terms of both accuracy and computational cost. We have carefully compared the performances of various carbon potentials in the LAMMPS code at our in-house 48 CPU sever. The results are summarized in Table S1. Clearly, the C-ADP potential is about a relatively same level of cost compared the widely used Tersoff potential. In addition to the traditional force fields (ADP, Tersoff, EDIP), we also employed the artificial neural network (ANN) potential in this work for the screening of low energy interface structures with small unit cell.

Though ANN is able to yield better accuracy (close to quantum mechanical simulation), it usually takes long time to in training to obtain the optimum ANN weight parameters. Moreover, the computational cost for each MD step is about 2-3 orders of magnitude higher than the traditional force field. Therefore, we did not consider ANN for MD simulation.

**Table S1.** Comparison of computational cost for different carbon potentials used in the LAMMPS code for 1600, 12800, 102400 carbon atoms.

| Potential | Cost (s/timestep with 48 CPUs) | | |
|---|---|---|---|
| | **1600 atoms** | **12800 atoms** | **102400 atoms** |
| Tersoff [5] | 0.034 | 0.159 | 0.432 |
| EDIP [6] | 0.203 | 0.512 | 6.472 |
| ANN [7] | 105.033 | 1082.308 | -- |
| C-ADP (this work) | 0.053 | 0.193 | 0.716 |

We compared the accuracy of ADP with DFT+D3 approach for the equations of states (EOS). The EOS results are summarized in Figure S1. In general, our ADP results agree with the DFT results very well. This indicates that ADP can yield reliable geometry comparable to DFT.

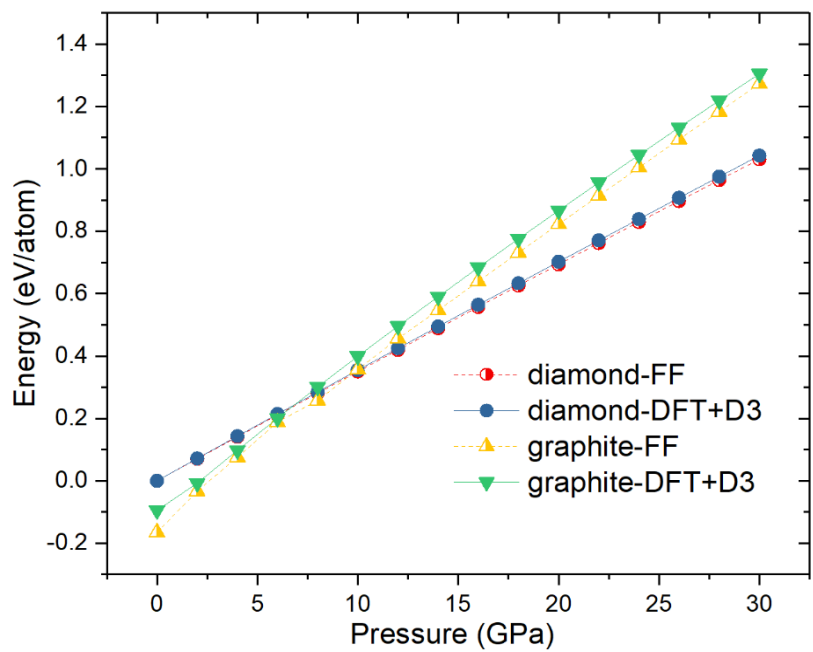

**Figure S1.** The equation of states of graphite and diamond calculated by DFT+D3 and FF (C-ADP in this work). The ambient pressure diamond energies are set to zero, both DFT+D3 and FF, respectively.

Further, we compared the accuracy of ADP with DFT+D3 approach for the energy barrier of hexagonal/cubic diamond to graphite transitions. The results are shown in Figure S2. Since there exist a systematic error between DFT and ADP results in the energy of graphite, the ADP barriers do not exactly fit the DFT barriers. However, the relative barriers are consistent. This seems unavoidable for any classical carbon potentials.

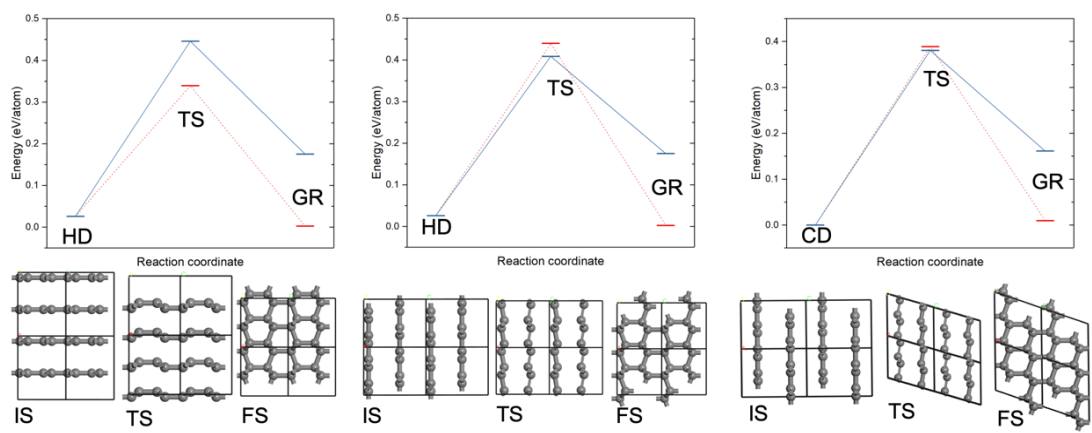

**Figure S2.** The comparison of energy barrier calculated by DFT (blue) and the ADP potential (red). The energy of cubic diamond is set to zero for comparison.

In sum, our ADP model generally yields very consistent geometries compared with the DFT results. But the energy is less accurate due to the limitation of EAM formulism. This trend is also observed in our simulation. For instance, the identified interface of t-(100)gr//(11-1)cd + [010]gr//[1-10]cd from our ADP-MD simulation agree with the experimental data very well. Moreover, the tilted graphite (001) plane in this interface will return to the ideal alignment under the geometry optimization in both DFT and force field calculations at 0 K. Based on the excellent agreement with the experiment, we believe the accuracy of ADP is sufficient for investigating the graphite-diamond transition at HPHT conditions.

## 3. High index phase propagation frontiers in graphite-diamond transition

In our MD simulation, we observed mainly the (100)gr//(11-1)cd interfaces which lead to {111} diamond stacking faults. However, it is also possible to form other interfaces with high index propagate frontiers. If the diamond nuclei propagate along [120]gr with different rates, different phase growth propagation frontiers will occur, as shown in Figure S3b. In this case, other high index twins or stacking faults may occur when these high index planes meet highly bent graphite during the phase propagation.

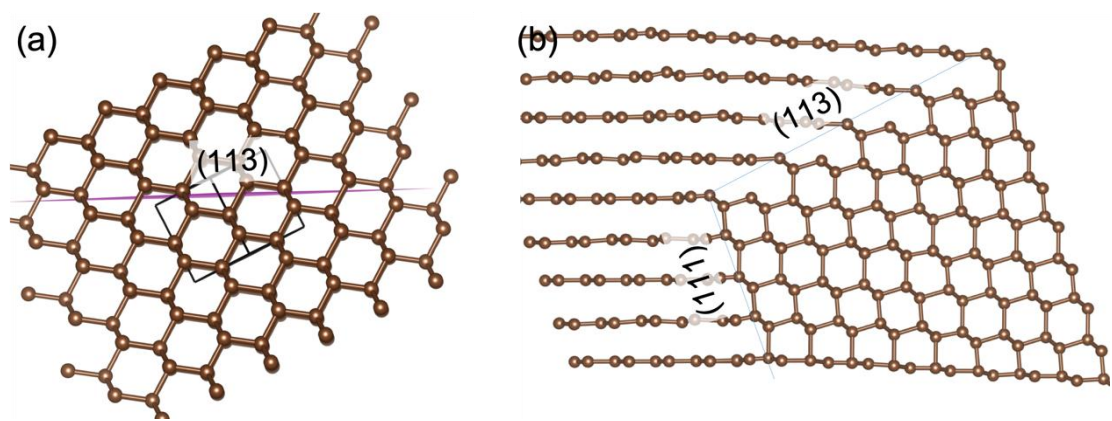

**Figure S3.** The schematic formation of diamond-(11-1) and (113) propagate frontiers by the lateral displacement mechanism. (a) is the diamond (113) plane in the ideal crystal. (b) is the phase growth propagate frontiers of (11-1) and (113) diamond.

## 4. Two directions {111} twins in previously reported TEM data

We found the two directions {111} twins were reported in several previous studies. They were shown in Figure S4. These results support our interpretation of the observation of twin structures in two different directions ([120]gr and [001]gr) very well. If the diamond nucleus only propagates along [001], only one (111) twin can occur.

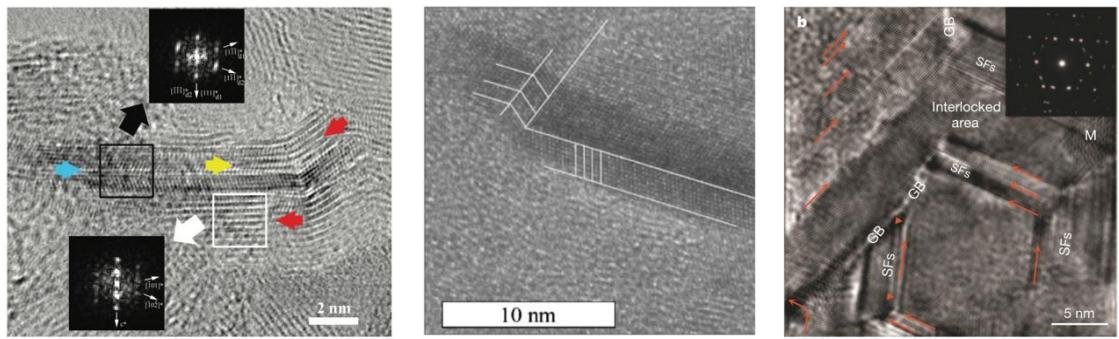

**Figure S4. The high resolution TEM of two direction (111) twinning structure.** (a) is from ref. [33] (b) is from ref. [29] (c) is from ref. [18].

Moreover, the graphite sheets adopt a purely planar configuration is much harder to produce the [120]gr direction twins, that's why many synthetic diamonds from single crystal graphite were reported to have the lamellar texture with only one direction of twin structures along [001]gr direction in a nano domain, see Figure S5.

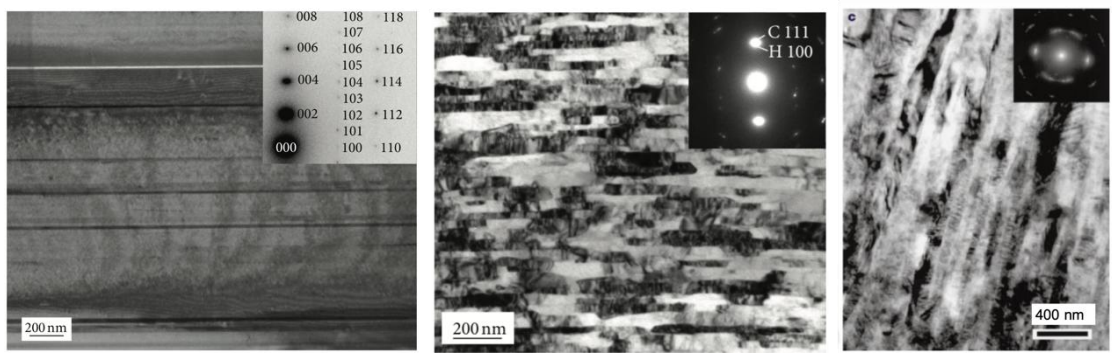

**Figure S5. The lamellar texture cubic diamond with up to 100-200 nm in length but several nanometers in thickness.** (a) and (b) is from ref. 5. (c) is from ref. 45.